\documentclass[english]{article}
\usepackage[T1]{fontenc}
\usepackage[latin9]{inputenc}
\usepackage{float}
\usepackage{amsmath}
\usepackage{graphicx}

\makeatletter

\providecommand{\tabularnewline}{\\}

\usepackage{braket}
 
\usepackage{fullpage}

\makeatother

\usepackage{babel}
\begin{document}

\thispagestyle{empty} \vskip 3em \begin{center} {{\bf \Large Quantum graph model for rovibrational states of protonated methane }}  \\[10mm]
{\bf \large J.~I. Rawlinson\footnote{email: jir25@damtp.cam.ac.uk}} \\[1pt] \vskip 1em {\it  Department of Applied Mathematics and Theoretical Physics,\\ University of Cambridge, \\ Wilberforce Road, Cambridge CB3 0WA, U.K.} \vspace{12mm}
\end{center}

\begin{abstract}
We extend the quantum graph model for the protonated methane $\left(CH_{5}^{+}\right)$
molecular ion, allowing for orientational degrees of freedom. This
enables us to compute $J>0$ rovibrational states, and we present
our results for $J=0,1,2,3$.
\end{abstract}

\section*{Introduction}

In \cite{fabri} it was proposed that the nuclear motion for the protonated
methane $\left(CH_{5}^{+}\right)$ system can be usefully approximated
by motion on a quantum graph. The authors computed the energies and
symmetry properties of the vibrational states, comparing the results
with more sophisticated quantum-chemical calculations. The agreement
is remarkable given the simplicity of the quantum graph model, in
which the relevant internal degrees of freedom are taken to be only
one-dimensional (to be compared with seven-dimensional \cite{carrington,fabri7}
and even twelve-dimensional calculations \cite{carrington12}). These
quantum graph calculations were, however, restricted to $J=0$ states
as orientational degrees of freedom were neglected.

In this paper we extend these calculations to the $J>0$ sector, allowing
us to explore the full rovibrational spectrum for the quantum graph
model. It is here that we see the real advantage of the quantum graph
approach: the drastic reduction in the number of degrees of freedom
allows us to compute states which have been inaccessible to higher
dimensional quantum-chemical calculations. We illustrate this by computing
the full low-energy spectrum for $J=3$ rovibrational states.

\section*{Quantum graph model}

Recall the $120$-vertex graph $\Gamma$ (see Figure $1$) introduced
in \cite{fabri} . Each point on the graph represents a possible molecular
shape for $CH_{5}^{+}$, with the vertices corresponding to the $120$
symmetry-equivalent energy minima on the potential energy surface
(PES) and the edges representing low-energy paths between them. The
minima have a $C_{s}$ point-group symmetry and can be thought of
as a $H_{2}$ unit sitting on top of a $CH_{3}^{+}$ tripod. Each
is connected to three other minima, with two different kinds of paths
occuring, indicated by the red and blue edges. The blue edges correspond
to an internal rotations of the $H_{2}$ relative to the tripod (this
motion takes the configuration through a $C_{s}$-symmetric saddle
point). The red edges correspond to a flip motion which exhanges a
pair of protons between the $H_{2}$ and $CH_{3}^{+}$ units (taking
the configuration through a $C_{2v}$-symmetric saddle point). These
paths are illustrated in Figure $2$.

We assume that, even at low energies, the molecule is not rigid but
is able to explore all of this graph by changing its shape. The molecule
also has rotational degrees of freedom. So the space $\mathcal{C}$
of all possible configurations of the molecule can be thought of as
the product space of the graph $\Gamma$ with the space of possible
orientations $SO(3)$: $\mathcal{C}\cong\Gamma\times SO(3)$. Our
strategy is to map the (very complex) quantum dynamics of $CH_{5}^{+}$
onto the motion of a particle confined to $\mathcal{C}$.

\begin{figure}[H]
\centering{}\includegraphics[scale=0.6]{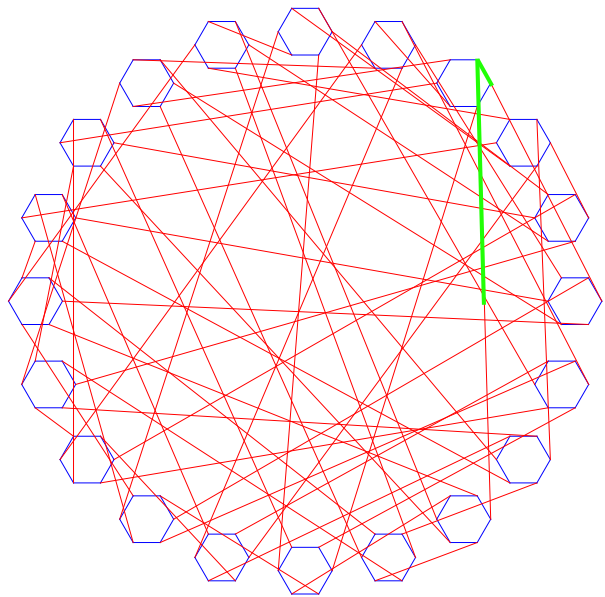}\caption{Quantum graph.}
\end{figure}

\begin{figure}[H]
\begin{centering}
\includegraphics[bb=0bp 175bp 695bp 724bp,scale=0.5]{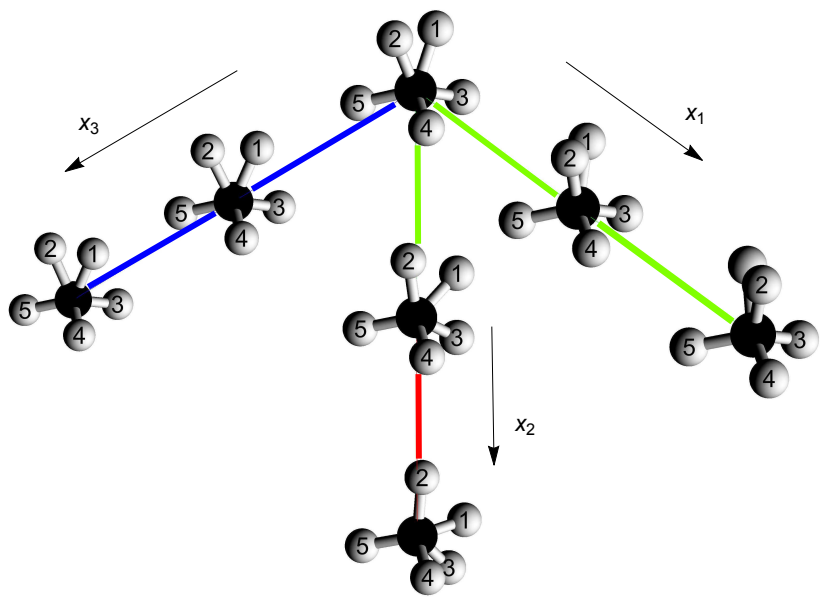}
\par\end{centering}
\caption{Low-energy paths between minima.}
\end{figure}

\subsection*{Motion on $\mathcal{C}$ and symmetries}

The motion of a particle confined to $\mathcal{C}$ is already a drastic
simplification compared to the full quantum dynamics of $CH_{5}^{+}$.
However, by exploiting symmetries, we can make the problem even simpler.
The Molecular Symmetry (MS) group of $CH_{5}^{+}$ is $S_{5}$, generated
by permutations of the $5$ protons. Each permutation $\pi\in S_{5}$
acts on $\mathcal{C}$, taking configurations at a given point $p$
on the graph $\Gamma$ and mapping them to configurations at a new
point $\pi\left(p\right)$ on the graph. In fact, by acting with elements
of $S_{5}$ we can generate the entire graph from only two edges,
or even one edge and one half-edge. An example of a choice is highlighted
in green in Figure $1$. We will refer to this green part of $\mathcal{C}$
as the \emph{fundamental domain} for $S_{5}$, and the vertex where
the two green edges meet as $V$. Note that $S_{5}$ is a symmetry
of $\mathcal{C}$ and so the quantum states can be classified by irreducible
representations (irreps) of $S_{5}$. Working within a particular
irrep, the wavefunctions on $\mathcal{C}$ must transform in a definite
way under the action of $S_{5}$ and this allows us to deduce the
value of the wavefunction on all of $\mathcal{C}$ so long as we know
the value of the wavefunction on the fundamental domain. So in fact
we only need to determine the wavefunction on the fundamental domain
(once an irrep has been chosen), not on all of $\mathcal{C}$: the
rest is determined by symmetry.

\subsection*{Defining the problem on the fundamental domain}

We need to write down an appropriate Hamiltonian on each edge. This
should involve a kinetic energy contribution and a potential. In order
to make the model as simple as possible we will set the potential
to zero. In general, one expects the kinetic energy to involve contributions
from both vibrational motion (motion along the graph) and rotational
motion as well as so-called rovibrational cross terms. But there is
quite a lot of freedom in which coordinates we choose and so we can
exploit this freedom to eliminate the cross terms: essentially, we
want vibrational motions to be orthogonal to rotational motions.

Start off by picking a coordinate $x'$ along the edge of the graph.
We will use Euler angles for the orientational degrees of freedom,
so altogether we have coordinates $(x',\alpha,\beta,\gamma)$. The
Euler angles $\left(\alpha,\beta,\gamma\right)$ tell us (in the usual
way) the rotation relating the body-fixed frame of the molecule to
a space-fixed frame. But we must still specify a choice of body-fixed
frame for each shape along the edge. It is clear that we can make
this choice, as we go along the edge, in such a way that compensates
for any angular momentum generated by the vibrational motion and thus
eliminates any cross terms. So we may assume that in the coordinates
$(x',\alpha,\beta,\gamma)$ the kinetic energy operator has only a
purely vibrational contribution and a purely rotational contribution.
Now we can further transform the coordinate $x'\rightarrow x\left(x'\right)$
so as to make the vibrational kinetic operator simply $-\frac{1}{2}\frac{d^{2}}{dx^{2}}$
(this relies on the fact that the vibration is only one-dimensional).
As for the rotational kinetic energy, we assume that the moments of
inertia do not vary much and so we use the approximate kinetic energy
operator $\frac{1}{2I}\hat{\mathbf{J}}^{2}$ where \textbf{$\hat{\mathbf{J}}$}
is the generator of body-fixed rotations and $I$ is a constant moment
of inertia. Thus we have
\begin{equation}
\mathcal{H}=-\frac{1}{2}\frac{d^{2}}{dx^{2}}+\frac{1}{2I}\hat{\mathbf{J}}^{2}.\label{eq:-1}
\end{equation}
We are now in a position to set up the problem on the fundamental
domain.

A vicinity of the fundamental domain is shown in Figure $2$, consisting
of the vertex $V$ together with the three edges leaving it. Let $x_{1},x_{3}\in\left[0,L_{1}\right]$
and $x_{2}\in\left[0,L_{2}\right]$ be coordinates along the edges
leaving the vertex, with the (green) fundamental domain corresponding
to $x_{1}\in\left[0,L_{1}\right]$ and $x_{2}\in\left[0,\frac{L_{2}}{2}\right]$.
Suppose $\Psi^{\left(T\right)}$ is a state which transforms in the
irrep $T$ of $S_{5}$ (so has degeneracy $\text{dim}T$) and that
for every $\pi\in S_{5}$ we have the corresponding matrix action
on the state
\begin{equation}
\Psi_{n}^{\left(T\right)}\rightarrow\sum_{m=1}^{\text{dim}T}T\left(\pi\right)_{nm}\Psi_{m}^{\left(T\right)}.\label{eq:-2}
\end{equation}
 On edge $j\in\left\{ 1,2,3\right\} $ the Hamiltonian is
\begin{equation}
\mathcal{H}_{j}=-\frac{1}{2}\frac{d^{2}}{dx_{j}^{2}}+\frac{1}{2I}\hat{\mathbf{J}}^{2}.\label{eq:-3}
\end{equation}
Recall from standard rigid-body theory that the rotational symmetry
implies states are classified by quantum numbers $J$ (total angular
momentum) and $M\in\left\{ -J,\ldots,+J\right\} $ (space-fixed angular
momentum projection). So we assume that $\Psi_{n}^{\left(T\right)}$
is a $\left(J,M\right)$ state. We can expand the wavefunction on
edge $j$ in terms of $\left(J,M\right)$ symmetric-top eigenfunctions
(with body-fixed angular momentum projection $K\in\left\{ -J,\ldots,+J\right\} $)
and plane waves:
\begin{equation}
\sum_{K=-J}^{J}\left(a_{njK}e^{ikx_{j}}+b_{njK}e^{ik\left(L_{j}-x_{j}\right)}\right)\ket{JKM}\label{eq:-4}
\end{equation}
with corresponding energy eigenvalues $E=\frac{1}{2}k^{2}+\frac{1}{2I}J\left(J+1\right)$.

Now recall that we have $S_{5}$ symmetry: for example, the permutation
$\left(12\right)\left(543\right)\in S_{5}$ maps configurations on
edge $1$ with $x_{1}=x$ to configurations on edge $3$ with $x_{3}=L_{1}-x$.
The orientations differ by some rotation $R=\exp\left(-i\theta\hat{\mathbf{n}}\cdot\hat{\mathbf{J}}\right)$
($\theta$ and $\hat{\mathbf{n}}$ are estimated in the appendix).
We can therefore deduce the wavefunction on edge $3$ from the wavefunction
on edge $1$. Explicitly, it is

\begin{equation}
\sum_{m=1}^{\text{dim\ensuremath{T}}}\sum_{K'=-J}^{J}\sum_{K=-J}^{J}T\left(\left(12\right)\left(543\right)\right)_{nm}\exp\left(-i\theta\hat{\mathbf{n}}\cdot\hat{\mathbf{J}}\right)_{KK'}\left(a_{m1K'}e^{ik\left(L_{1}-x_{3}\right)}+b_{m1K'}e^{ikx_{3}}\right)\ket{JKM}.\label{eq:-5}
\end{equation}
Now we impose the quantum graph boundary conditions at the vertex
$V$ joining edges $1,2$ and $3$. These are continuity of the wavefunction
together with current conservation, as explained in \cite{fabri}.
As we have expressed the wavefunction on edge $3$ in terms of its
values on edges $1$ and $2$, these boundary conditions give us some
new conditions relating just the wavefunctions on edges $1$ and $2$
which have to be satisfied. For example, continuity of the wavefunction
at $V$ ($x_{1}=x_{2}=x_{3}=0$) leads to
\begin{eqnarray}
\sum_{m=1}^{\text{dim\ensuremath{T}}}\sum_{K'=-J}^{J}T\left(\left(12\right)\left(543\right)\right)_{nm}\exp\left(-i\theta\hat{\mathbf{n}}\cdot\hat{\mathbf{J}}\right)_{KK'}\left(a_{m1K'}e^{ikL_{1}}+b_{m1K'}\right) & = & \left(a_{n1K}+b_{n1K}e^{ikL_{1}}\right)\label{eq:}\\
 & = & \left(a_{n2K}+b_{n2K}e^{ikL_{2}}\right)\nonumber 
\end{eqnarray}

A similar calculation, considering the permutation $\left(23\right)\left(45\right)$,
gives boundary conditions at the midpoint of edge $2$. Thus we end
up with a set of linear equations in the variables $a_{n1K},a_{n2K},b_{n1K},b_{n2K}$
which, as we see in (\ref{eq:}), depend on momentum $k$. These are
our quantization conditions, and can be handled numerically.

We should note that there is an additional symmetry present, namely
inversion in space, which can be used to classifiy states in addition
to $S_{5}$. We have found the parity of our computed states by noting
that, on edge $2$, spatial inversion can be realised by the combined
action of the permutation $\left(45\right)\in S_{5}$ followed by
a rotation by $\pi$ about the axis normal to the plane of $C_{s}$
reflection symmetry.

\section*{Results and discussion}

We display the lowest-lying rovibrational states in Tables $1$-$4$,
listed against reference data from the 7-dimensional variational calculation
in \cite{carrington}. We have used the values $L_{1}=61.2\sqrt{m_{e}}a_{0}$
and $L_{2}=1.0\sqrt{m_{e}}a_{0}$, following the suggestion in \cite{fabri},
to give the best fit to the $J=0$ data (here $m_{e}$ is the electron
mass and $a_{0}$ the Bohr radius). For the moment of inertia we have
picked a physically reasonable value $\frac{1}{I}=8$ $\text{c}\text{m}^{-1}$.

We see that the quantum graph states give a good qualitative fit to
the reference data even when we extend to $J>0$, with correct $S_{5}$
irrep and parity assignments along with reasonable energy values.
The agreement is remarkable considering the simplicity of the quantum
graph model. We expect the model to break down at higher energies,
where neglected degrees of freedom become important, but these results
demonstrate that the graph model is sufficient to understand many
states in the low-energy regime. Our $J=3$ states go beyond those
computed in \cite{carrington}, in which full spectra were only calculated
for $J\leq2$. Based on the agreement in the $J\leq2$ sector, we
expect our $J=3$ states to be a reliable description of the states
of $CH_{5}^{+}$ in the energy range considered.

\begin{table}[H]
\begin{centering}
\begin{tabular}{|ccc|ccc|}
\hline 
Irrep & $E$ & $E_{\text{ref}}$ & Irrep & $E$ & $E_{\text{ref}}$\tabularnewline
\hline 
$A_{1}^{+}$ & $0.0$ & $0.0$ & $G_{2}^{-}$ & $11.4$ & $9.9$\tabularnewline
$H_{1}^{+}$ & $22.2$ & $20.4$ & $H_{2}^{-}$ & $39.6$ & $41.1$\tabularnewline
$G_{1}^{+}$ & $44.8$ & $49.4$ & $I^{-}$ & $49.7$ & $58.4$\tabularnewline
$H_{2}^{+}$ & $50.2$ & $59.3$ & $H_{1}^{-}$ & $96.0$ & $113.7$\tabularnewline
$I^{+}$ & $95.2$ & $112.0$ & $G_{2}^{-}$ & $100.9$ & $112.7$\tabularnewline
$H_{1}^{+}$ & $112.4$ & $122.0$ & $H_{2}^{-}$ & $148.8$ & $139.4$\tabularnewline
\hline 
\end{tabular}
\par\end{centering}
\caption{$J=0$ States for quantum graph model (exactly reproducing the results
in \cite{fabri}). The reference data $E_{\text{ref}}$ is from \cite{carrington}.}
\end{table}

\begin{table}[H]
\begin{centering}
\begin{tabular}{|ccc|ccc|}
\hline 
Irrep & $E$ & $E_{\text{ref}}$ & Irrep & $E$ & $E_{\text{ref}}$\tabularnewline
\hline 
$I^{+}$ & $15.3$ & $14.7$ & $G_{1}^{-}$ & $11.6$ & $11.3$\tabularnewline
$G_{2}^{+}$ & $25.5$ & $23.0$ & $H_{2}^{-}$ & $27.3$ & $24.9$\tabularnewline
$H_{2}^{+}$ & $32.5$ & $31.9$ & $I^{-}$ & $30.9$ & $29.7$\tabularnewline
$I^{+}$ & $43.9$ & $46.5$ & $H_{1}^{-}$ & $38.4$ & $40.3$\tabularnewline
$G_{2}^{+}$ & $51.4$ & $57.9$ & $A_{1}^{-}$ & $49.9$ & $55.4$\tabularnewline
$H_{1}^{+}$ & $53.4$ & $57.1$ & $G_{2}^{-}$ & $52.6$ & $62.1$\tabularnewline
$G_{1}^{+}$ & $54.1$ & $61.3$ & $I^{-}$ & $55.3$ & $59.6$\tabularnewline
$H_{2}^{+}$ & $62.9$ & $72.0$ & $H_{1}^{-}$ & $57.3$ & $67.1$\tabularnewline
$I^{+}$ & $64.1$ & $72.1$ & $G_{1}^{-}$ & $58.8$ & $62.3$\tabularnewline
$H_{1}^{+}$ & $92.7$ & $115.1$ & $H_{2}^{-}$ & $66.6$ & $75.5$\tabularnewline
$G_{2}^{+}$ & $101.4$ & $117.1$ & $I^{-}$ & $94.9$ & $115.0$\tabularnewline
$H_{2}^{+}$ & $103.1$ & $117.2$ & $H_{1}^{-}$ & $96.8$ & $115.9$\tabularnewline
$I^{+}$ & $107.2$ & $122.7$ & $G_{2}^{-}$ & $107.8$ & $122.5$\tabularnewline
$G_{1}^{+}$ & $112.3$ & $126.5$ & $H_{2}^{-}$ & $109.9$ & $122.6$\tabularnewline
$A_{2}^{+}$ & $113.5$ & $125.8$ & $G_{1}^{-}$ & $113.8$ & $126.3$\tabularnewline
$H_{2}^{+}$ & $137.0$ & $138.7$ & $I^{-}$ & $127.1$ & $134.0$\tabularnewline
$I^{+}$ & $154.2$ & $145.2$ & $H_{1}^{-}$ & $149.9$ & $143.4$\tabularnewline
\hline 
\end{tabular}
\par\end{centering}
\caption{$J=1$ States for quantum graph model.}

\end{table}
\begin{table}[H]
\begin{centering}
\begin{tabular}{|ccc|ccc|ccc|ccc|}
\hline 
Irrep & $E$ & $E_{\text{ref}}$ & Irrep & $E$ & $E_{\text{ref}}$ & Irrep & $E$ & $E_{\text{ref}}$ & Irrep & $E$ & $E_{\text{ref}}$\tabularnewline
\hline 
$H_{1}^{+}$ & $29.5$ & $29.1$ & $G_{2}^{+}$ & $88.5$ & $98.5$ & $H_{2}^{-}$ & $32.1$ & $31.6$ & $G_{1}^{-}$ & $87.8$ & $92.0$\tabularnewline
$G_{1}^{+}$ & $32.2$ & $32.2$ & $A_{1}^{+}$ & $90.4$ & $92.2$ & $I^{-}$ & $34.2$ & $33.2$ & $A_{2}^{-}$ & $99.7$ & $126.2$\tabularnewline
$H_{2}^{+}$ & $40.7$ & $39.5$ & $G_{1}^{+}$ & $102.1$ & $126.1$ & $H_{1}^{-}$ & $36.4$ & $36.9$ & $G_{1}^{-}$ & $101.6$ & $126.8$\tabularnewline
$H_{1}^{+}$ & $46.3$ & $46.7$ & $G_{2}^{+}$ & $102.6$ & $127.9$ & $A_{2}^{-}$ & $50.9$ & $45.0$ & $H_{2}^{-}$ & $103.4$ & $126.3$\tabularnewline
$G_{2}^{+}$ & $52.7$ & $48.5$ & $I^{+}$ & $105.7$ & $126.5$ & $G_{2}^{-}$ & $52.2$ & $50.0$ & $I^{-}$ & $108.0$ & $128.0$\tabularnewline
$I^{+}$ & $53.4$ & $53.9$ & $H_{1}^{+}$ & $111.5$ & $129.8$ & $I^{-}$ & $53.3$ & $54.6$ & $H_{2}^{-}$ & $117.6$ & $132.9$\tabularnewline
$H_{2}^{+}$ & $54.8$ & $59.3$ & $H_{2}^{+}$ & $115.4$ & $134.7$ & $H_{2}^{-}$ & $55.3$ & $56.4$ & $G_{2}^{-}$ & $118.2$ & $133.3$\tabularnewline
$G_{1}^{+}$ & $55.8$ & $59.2$ & $I^{+}$ & $127.2$ & $138.8$ & $G_{2}^{-}$ & $58.1$ & $64.8$ & $H_{1}^{-}$ & $121.1$ & $137.6$\tabularnewline
$A_{2}^{+}$ & $62.0$ & $73.3$ & $H_{2}^{+}$ & $128.7$ & $140.9$ & $G_{1}^{-}$ & $60.8$ & $66.1$ & $I^{-}$ & $127.6$ & $141.3$\tabularnewline
$I^{+}$ & $68.5$ & $76.5$ & $G_{2}^{+}$ & $140.1$ & $147.6$ & $H_{1}^{-}$ & $67.4$ & $73.4$ & $A_{1}^{-}$ & $139.9$ & $150.6$\tabularnewline
$H_{2}^{+}$ & $72.5$ & $78.9$ & $G_{1}^{+}$ & $146.8$ & $152.7$ & $I^{-}$ & $68.8$ & $74.5$ & $G_{2}^{-}$ & $148.8$ & $152.0$\tabularnewline
$H_{1}^{+}$ & $74.3$ & $82.6$ & $H_{1}^{+}$ & $151.3$ & $151.6$ & $H_{1}^{-}$ & $76.9$ & $82.5$ & $H_{1}^{-}$ & $152.4$ & $155.2$\tabularnewline
$G_{1}^{+}$ & $74.9$ & $78.1$ & $I^{+}$ & $154.6$ & $154.3$ & $H_{2}^{-}$ & $79.8$ & $86.7$ & $I^{-}$ & $155.1$ & $154.8$\tabularnewline
$I^{+}$ & $82.5$ & $88.5$ & $G_{1}^{+}$ & $158.0$ & $157.0$ & $I^{-}$ & $84.0$ & $94.4$ & $G_{1}^{-}$ & $157.8$ & $152.4$\tabularnewline
$H_{1}^{+}$ & $86.8$ & $95.0$ & $A_{1}^{+}$ & $161.8$ & $148.4$ & $G_{2}^{-}$ & $87.7$ & $96.4$ & $H_{2}^{-}$ & $171.1$ & $162.3$\tabularnewline
\hline 
\end{tabular}
\par\end{centering}
\caption{$J=2$ States for quantum graph model.}
\end{table}

\begin{table}[H]
\begin{centering}
\begin{tabular}{|cc|cc|cc|cc|}
\hline 
Irrep & $E$ & Irrep & $E$ & Irrep & $E$ & Irrep & $E$\tabularnewline
\hline 
$I^{+}$ & $56.1$ & $G_{1}^{+}$ & $113.1$ & $H_{1}^{-}$ & $54.6$ & $A_{2}^{-}$ & $115.9$\tabularnewline
$H_{1}^{+}$ & $59.0$ & $H_{1}^{+}$ & $115.3$ & $G_{1}^{-}$ & $56.4$ & $I^{-}$ & $115.9$\tabularnewline
$G_{2}^{+}$ & $61.6$ & $G_{2}^{+}$ & $118.6$ & $I^{-}$ & $59.2$ & $H_{2}^{-}$ & $117.7$\tabularnewline
$H_{2}^{+}$ & $61.9$ & $I^{+}$ & $119.8$ & $G_{2}^{-}$ & $66.9$ & $I^{-}$ & $117.8$\tabularnewline
$I^{+}$ & $63.0$ & $H_{1}^{+}$ & $120.2$ & $H_{1}^{-}$ & $67.2$ & $G_{1}^{-}$ & $118.7$\tabularnewline
$G_{1}^{+}$ & $68.1$ & $A_{1}^{+}$ & $121.1$ & $A_{1}^{-}$ & $68.0$ & $H_{2}^{-}$ & $122.8$\tabularnewline
$H_{2}^{+}$ & $78.4$ & $I^{+}$ & $122.1$ & $I^{-}$ & $68.7$ & $G_{2}^{-}$ & $124.4$\tabularnewline
$G_{2}^{+}$ & $79.3$ & $H_{2}^{+}$ & $125.6$ & $H_{2}^{-}$ & $74.6$ & $H_{1}^{-}$ & $124.8$\tabularnewline
$I^{+}$ & $80.4$ & $H_{1}^{+}$ & $131.9$ & $H_{1}^{-}$ & $79.3$ & $G_{1}^{-}$ & $125.4$\tabularnewline
$A_{2}^{+}$ & $82.0$ & $G_{2}^{+}$ & $140.0$ & $G_{1}^{-}$ & $79.7$ & $I^{-}$ & $135.7$\tabularnewline
$I^{+}$ & $83.3$ & $H_{2}^{+}$ & $143.3$ & $H_{2}^{-}$ & $81.5$ & $H_{2}^{-}$ & $140.6$\tabularnewline
$G_{1}^{+}$ & $88.1$ & $I^{+}$ & $143.9$ & $G_{2}^{-}$ & $81.6$ & $H_{1}^{-}$ & $152.5$\tabularnewline
$G_{2}^{+}$ & $89.6$ & $I^{+}$ & $151.4$ & $G_{2}^{-}$ & $90.0$ & $H_{2}^{-}$ & $154.3$\tabularnewline
$H_{1}^{+}$ & $90.0$ & $G_{1}^{+}$ & $156.4$ & $I^{-}$ & $92.8$ & $G_{1}^{-}$ & $156.6$\tabularnewline
$H_{2}^{+}$ & $92.3$ & $G_{1}^{+}$ & $171.4$ & $I^{-}$ & $97.5$ & $G_{2}^{-}$ & $157.7$\tabularnewline
$H_{1}^{+}$ & $99.5$ & $H_{1}^{+}$ & $171.4$ & $H_{1}^{-}$ & $98.0$ & $A_{1}^{-}$ & $170.8$\tabularnewline
$I^{+}$ & $103.8$ & $G_{2}^{+}$ & $173.6$ & $G_{1}^{-}$ & $99.7$ & $G_{1}^{-}$ & $176.0$\tabularnewline
$G_{2}^{+}$ & $111.7$ & $H_{2}^{+}$ & $174.5$ & $H_{2}^{-}$ & $106.0$ & $H_{1}^{-}$ & $176.8$\tabularnewline
$H_{2}^{+}$ & $112.7$ & $H_{1}^{+}$ & $185.1$ & $H_{1}^{-}$ & $112.3$ &  & \tabularnewline
\hline 
\end{tabular}
\par\end{centering}
\caption{$J=3$ States for quantum graph model.}
\end{table}

\section*{Appendix}

\subsection*{Estimating $R$}

A reasonable approximation is to take the positions of the protons
to be on the surface of a sphere (centred on the Carbon nucleus) as
illustrated below. We take the bond angles to be those which give
the closest match of the proton positions to \emph{ab initio }values:
the polar angle of the $H_{2}$ unit (in radians) is taken to be $0.42$
while the polar angle of the other three protons is taken to be $1.89$.

Consider configurations along edge $1$. Our choice of body-fixed
axes ($x,y,z$) are indicated in the picture: notice that as the $H_{2}$
unit rotates relative to the $CH_{3}$ tripod, the entire molecule
also rotates at a rate such that the total angular momentum vanishes.
The vanishing of the angular momentum ensures that, for this choice
of body-fixed axes, there are no kinetic rotation-vibration cross-terms.
The rates of rotation are related by the ratio of the moments of inertia,
$I_{1}$ and $I_{2}$, of the $H_{2}$ and of the whole molecule.
In particular, by the point that the $H_{2}$ unit has rotated a full
$\frac{2\pi}{6}$ with respect to the $CH_{3}$ tripod, the molecule
as a whole will have rotated in the opposite sense by $\Delta\theta=\left(\frac{I_{1}}{I_{2}}\right)\left(\frac{2\pi}{6}\right).$
Then the rotation relating the configurations at the two endpoints
of edge $1$ is a rotation by $\frac{2\pi}{3}+\Delta\theta\approx2.21$
about the body-fixed $z$-axis, and so we take 
\[
R=\exp\left(-i\theta\hat{\mathbf{n}}\cdot\hat{\mathbf{J}}\right)
\]
 with 
\[
\theta\approx2.21,\hat{\mathbf{n}}=\begin{pmatrix}0\\
0\\
-1
\end{pmatrix}
\]
 in our boundary conditions. Edge $2$ is treated similarly.
\begin{center}
\begin{figure}[H]
\centering{}\includegraphics[scale=0.5]{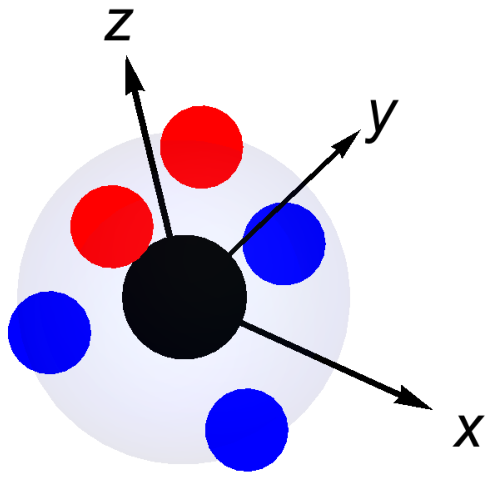}\includegraphics[scale=0.5]{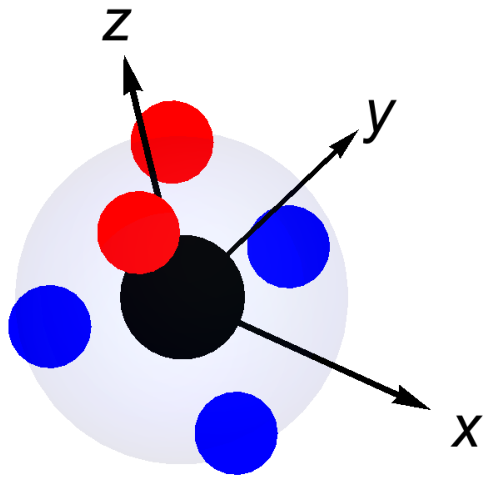}\includegraphics[scale=0.5]{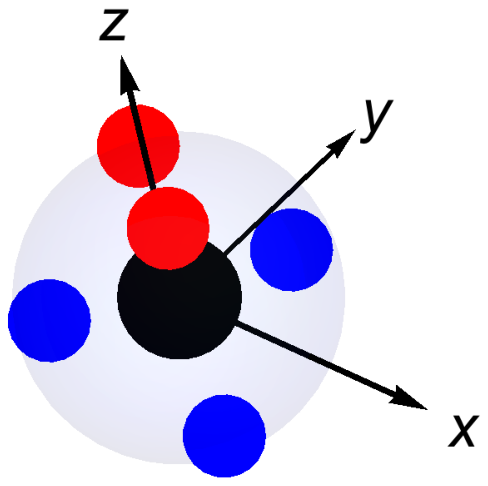}\caption{Choice of body-fixed axes.}
\end{figure}
\par\end{center}

\section*{Acknowledgements}

I am grateful to Csaba Fabri and Attila G. Csaszar for providing Figure
$1$ and for useful comments. I am supported by an EPSRC studentship.
This work has been partially supported by STFC consolidated grant
ST/P000681/1.

\end{document}